\documentclass[a4paper,11pt]{article}
\usepackage{latexsym}
\usepackage{amsmath}
\usepackage{amssymb}
\usepackage{bm}
\title{ Mixed tensor susceptibility of the QCD vacuum from
effective quark-quark interactions}
\author{$\mbox{Zhao Zhang}^{a,1}$, $\mbox{Wei-qin
Zhao}^{b,a}$\\[5pt] \textit{${}^a$Institute of High Energy
Physics, Academica Sinica,}\\ \textit{Beijing, 100039, P. R.
China}\\
 \textit{${}^b$CCAST(World Laboratory), P.O. Box 8730,}\\
\textit{Beijing 100080, P. R. China}}
\date{}
\begin{document}
\maketitle
 \footnotetext[1]{E-mail: zhangzh@ccastb.ccast.ac.cn }
\begin{abstract}
We calculate the mixed tensor susceptibility of QCD vacuum  in the
framework of the global color symmetry model. In our calculation,
the functional integration over gluon fields can be performed and
the gluonic vacuum observable can be expressed in terms of the
quark operators and the gluon propagator. The mixed tensor
susceptibility was obtained with the subtraction of the
perturbative contribution which is evaluated by the  Wigner
solution of the quark gap equation. Using several different
effective quark-quark interaction models, we find the values of
the mixed tensor susceptibility are very small.\\
\\[4pt] {\textit{Keywords}}:\, QCD vacuum ; Induced
vacuum condensate; Mixed tensor susceptibility ;Global color
symmetry model; Dyson-Schwinger equations\\
PACS number(s): 11.15.Tk; 12.38.Aw; 12.38.Lg; 12.39.-x
\end{abstract}
\newpage
\noindent{\textbf{\Large{1. Introduction}}} \vspace{5pt}

In SVZ sum rules, in order to determine the static properties of
hadrons it was suggested to consider two-point correlator
functions of quark currents in the presence of an external
constant classical field, where nonperturbative effects are taken
into account in the so-called vacuum susceptibilities\cite{r1}.
These induced condensates play important roles in determination of
the hadron properties such as the nucleon magnetic
moments\cite{r1}, the isovector axial coupling
constant\cite{r2,r3,r4}, the isoscalar axial coupling
constant\cite{r4}, the pion-nucleon coupling constant\cite{r5} and
the nucleon tensor charge\cite{r6,r7} within this version of SVZ
sum rules. In the literature, there are always two kinds of vacuum
susceptibility that appear in the conventional two-point treatment
of an external current field : one is the induced quark condensate
and the other is the induced mixed quark-gluon condensate. For
convenience, we refer to the former as the quark condensate
susceptibility and the later as mixed condensate susceptibility in
this letter.

The vacuum tensor susceptibilities are relevant for the
determination of nucleon tensor charge\cite{r6,r7}. The value of
nucleon charge is related to the first moment of the transversity
distribution $h_1(x)$, where $h_1(x)$ is an additional twist-two
chirality violating structure function which can be measured in
the Drell-Yan process with both beam and target transversely
polarized. The previous evaluation of the quark condensate tensor
susceptibility were performed in the framework of QCD sum
rules\cite{r6,r7,r8,r9}, the chiral constituent model\cite{r10}
and global color symmetry model(GCM)\cite{r11} respectively.
Actually, there still exist uncertainty about this induced
susceptibility since different theoretical treatments can give
very different results,  which should be checked by the future
measurement of the transversity distribution $h_1(x)$. Another
tensor susceptibility, the mixed condensate tensor susceptibility
was only evaluated roughly in Ref\cite{r6}. within the two-point
function of  QCD sum rules. In this letter, we will give the
calculation of the vacuum mixed tensor susceptibility within the
framework of GCM in the mean field level.

 As a truncated DSE-model, GCM is a quite successful
four-fermion interaction field theory which can be directly
derived through a truncation of QCD\cite{r13,r14}. This truncated
DSE model has been applied extensively at zero temperature and
chemical potential to the phenomenology of QCD\cite{r15,r16},
including the studies of observables from strong interaction to
weak interaction area. Furthermore, the truncated DSE models also
have made important progress in the studies of strong QCD at
finite temperature and chemical potential\cite{r17}. Due to the
fact that the evaluation of mixed quark-gluon condensate can be
performed in the framework of GCM\cite{r12}, a method to evaluate
the mixed vacuum tensor susceptibility is proposed within this DSE
formalism  and the numerical results for the mixed tensor
susceptibility are given.
\par

\vspace{5pt} \noindent{\textbf{\Large{2. Formalism}}} \vspace{5pt}

In the chiral limit, the QCD generating functional for quark field
in the Euclidean space is given by\\
\begin{equation}
\mathcal{Z}[\overline\eta,\eta]=\int\mathcal{D}q\mathcal{D}\overline{q}
\mathcal{D}\overline{w}\mathcal{D}w\mathcal{D}A\exp
\left\{-S-S_{gf}-S_g+\int{d}^4x(\overline\eta{q}+\overline{q}\eta)\right\},
\end{equation}
where
\begin{equation}
S=\int{d}^4x\left\{\overline{q}\left[\gamma_\mu(\partial_\mu-i{g}\frac{\lambda^a}{2}A^a_\mu)
\right]q+\frac{1}{4}F^a_{\mu\nu}F^a_{\mu\nu} \right\},
\end{equation}
and $S_{gf}$, $S_g$ are  the gauge-fixing and ghost actions
respectively.  Through introducing a bilocal field $B^\theta(x,y)$
as in \cite{r8,r9,r10}, the generating functional can be given as
\begin{eqnarray}
\mathcal{Z}[\overline{\eta},
\eta]&=&\exp\left[W_1(i{g}\frac{\delta}{\delta\eta(x)}\frac{\lambda^a}{2}\gamma_\mu
\frac{\delta}{\delta\overline{\eta}(x)})
\right]\int\mathcal{D}\overline{q}\mathcal{D}q
\mathcal{D}B^\theta(x, y)\nonumber\\
&&\times\exp\left\{-S[\overline{q}, q, B^\theta(x,
y)]+\int{d}^4x(\overline{\eta}q +\overline{q}\eta)\right\},
\end{eqnarray}
where
\begin{equation}
W_1[J^a_\mu]=\sum_{n=3}^{\infty}\frac{1}{n!}\int{d}^4x_1\cdots{d}^4x_nD^{a_1\cdots{a_n}}
_{\mu_1\cdots\mu_n}(x_1,\cdots,x_n)\prod^n_{i=1}J^{a_i}_{\mu_i}(x_i),
\end{equation}
and

\begin{eqnarray}
S[\overline{q},q,B^\theta(x,y)]&=&\int\int{d}^4xd^4y\left[\overline{q}(x)G^{-1}(x,y;[B^\theta])
q(y)\right.\nonumber\\
&&+\left.\frac{B^\theta(x,y)B^\theta(y,x)}{2g^2D(x-y)}\right],
\end{eqnarray}
with
\begin{equation}
G^{-1}(x,y;[B^\theta])=\gamma\cdot\partial\delta(x-y)+\Lambda^\theta{B}^\theta(x,y).
\end{equation}
The quantity $\Lambda^\theta$ arises from Fierz reordering
transformation and is the direct product of Dirac, flavor SU(3)
and color matrices
\begin{equation}
\Lambda^\theta=\frac{1}{2}(1_D,i\gamma_5,\frac{i}{\sqrt{2}}\gamma_\nu,
\frac{i}{\sqrt{2}}\gamma_\nu\gamma_5)\otimes(\frac{1}{\sqrt{3}}1_F,\frac{1}{\sqrt{2}}\lambda^a_F)\otimes
(\frac{4}{3}1_C,\frac{i}{\sqrt{3}}\lambda^a_C).
\end{equation}
And $g^2D(x-y)$ is the connected  gluon two-point function without
quark-loop contributions.
\par
Neglecting $W_1[J^a_\mu]$, we can obtain the GCM generating
functional as
\begin{equation}
\mathcal{Z}_{GCM}[\overline{\eta},\eta]=\int\mathcal{D}\overline{q}\mathcal{D}q\mathcal\mathcal{D}{B^\theta}(x,y)
\exp\left\{-S[\overline{q},q,B^\theta(x,y)]+\int{d}^4x(\overline{\eta}q+\overline{q}\eta)\right\}.
\end{equation}
Performing the functional integration over
$\mathcal{D}\overline{q}$ and $\mathcal{D}q$ in above equation, we
obtain
\begin{equation}\label{qf}
\mathcal{Z}_{GCM}[\overline{\eta},\eta]=\int\mathcal{D}B^\theta(x,y)\exp(-S[\overline{\eta},\eta
,B^\theta(x,y)]),
\end{equation}
where
\begin{eqnarray}
S[\overline{\eta},\eta,B^\theta(x,y)]&=&-Tr\ln\left[\partial\cdot\gamma\delta(x-y)+
\Lambda^\theta{B}^\theta(x,y)\right]\nonumber\\
&&+\int\int\left[\frac{B^\theta(x,y)B^\theta(y,x)}{2g^2D(x-y)}+\overline{\eta}(x)
G(x,y;B^\theta)\eta(y)\right].
\end{eqnarray}
The saddle point of this action is defined as
$\delta{S}[\overline{\eta},\eta,B^\theta(x,y)]/
\delta{B}^\theta(x,y)|_{\overline{\eta}=\eta=0}=0$
 and is given
by
\begin{equation}\label{DSE}
B^\theta_0(x-y)=g^2D(x-y)tr_{\gamma{C}}[\Lambda^\theta{G}_0(x-y)].
\end{equation}
where $G_0$ stands for $G[B^\theta_0]$ and the trace is to be
taken in Dirac and color space, whereas the flavor trace has been
separated out.
\par
We will calculate the induced QCD vacuum condensates from the
saddle-point expansion, that is, we will work at the mean field
level. This is consistent with the large $N_c$ limit in the quark
fields for a given model gluon two-point function. Under the mean
field approximation, the dressed quark propagator $G(x-y)$ is
substituted by $G_0(x-y)$ which has the decomposition
\begin{equation}
G_0^{-1}(p)=i\gamma\cdot{p}+\Sigma(p)=i\gamma\cdot{p}A(p^2)+B(p^2).
\end{equation}
The $\Sigma(p)$ stands for the  dressing self-energy  of quarks
and is defined as
\begin{eqnarray}
\Sigma(p)=\Lambda^\theta{B}^\theta_0(p)=i\gamma\cdot{p}[A(p^2)-1]+B(p^2),
\end{eqnarray}
where the self-energy functions $A(p^2)$ and $B(p^2)$ are
determined by the rainbow Dyson-Schwinger equations (DSEs)
\begin{eqnarray}\label{dsea}
[A(p^2)-1]p^2&=&\frac{8}{3}\int\frac{d^4q}{(2\pi)^4}g^2D(p-q)
\frac{A(q^2)p\cdot{q}}{q^2A^2(q^2)+B^2(q^2)},\\\label{dseb}
B(p^2)&=&\frac{16}{3}\int\frac{d^4q}{(2\pi)^4}g^2D(p-q)
\frac{B(q^2)}{q^2A^2(q^2)+B^2(q^2)}.
\end{eqnarray}
Because the form of the gluon propagator $g^2D(s)$ in the
infrared region is unknown, one often uses various model forms
\cite{r15,r16,r17} as input parameters in the previous studies of
the Rainbow DSE.
\par
 Within the DSE formalism, there are two qualitatively
distinct solutions in Eq(\ref{dsea}) and (\ref{dseb}). The
``Nambu-Goldstone" solution characterized by $B(p^2)\neq{0}$
describes a phase : $(1)$ chiral symmetry is dynamically broken
for it provides a momentum dependent constituent quark mass
$M(p^2)=B(p^2)/A(p^2)$; and $(2)$ the dressed quarks are confined
for the dressed quark propagator does not have a Lehmann
representation\cite{r16}.  The alternative ``Wigner" solution
characterized by $B(p^2)\equiv{0}$ describes a phase with neither
dynamical chiral symmetry breaking nor confinement. In this
letter, we refer to the quark propagator in terms of the trivial
solution of the gap equation as the ``perturbative" quark
propagator $G^{per}_0(p^2)$ with only the vector part
\begin{equation}
[A'(p^2)-1]p^2=\frac{8}{3}\int\frac{d^4q}{(2\pi)^4}g^2D(p-q)
\frac{p\cdot{q}}{q^2A'(q^2))}.
\end{equation}
This ``perturbative" quark propagator can be seen as the expectation value of the operator $T[q_i(x)\overline{q}_j(y)]$ over
the perturbative vacuum $|P\rangle$ at the mean field level in the framework of GCM.

\par
From the GCM generating functional, it is now straightforward to
calculate the vacuum expectation value(VEV) of any quark operator
of the form
\begin{equation}\label{qop}
\mathcal{O}_n\equiv(\overline{q}_{j_1}\Lambda^{(1)}_{j_1i_1}q_{i_1})
(\overline{q}_{j_2}\Lambda^{(2)}_{j_2i_2}q_{i_2})\cdots
(\overline{q}_{j_n}\Lambda^{(n)}_{j_ni_n}q_{i_n}) ,
\end{equation}
in the mean field vacuum. Here the $\Lambda^{(i)}$ stands for an
operator in Dirac, flavor, and color space. Take the appropriate
number of derivatives with respect to external source terms
$\eta_i$ and $\overline\eta_j$ of Eq. (\ref{qf}) and set
$\eta_i=\overline\eta_j=0$ \cite{r21}, we can get
\begin{eqnarray}\label{vev}
\langle\mathcal{O}_n\rangle=(-1)^n\sum_p(-)^p[\Lambda^{(1)}_{j_1i_1}\cdots
\Lambda^{(n)}_{j_ni_n}(G_0)_{i_1j_{p(1)}}\cdots(G_0)_{i_nj_{p(n)}}],
\end{eqnarray}
where $p$ stands for a permutation of the $n$ indices. Once the
dressing quark propagator $G_0(q^2)$ (We ignore the subscript $0$
below)is determined, one can calculate the two quark condensate
$\langle\overline{q}q\rangle$ , the four quark
condensate$\langle\overline{q}\Lambda^{(1)}q\overline{q}\Lambda^{(2)}q\rangle$
, etc. in the mean field level. Since the functional integration
over the gluon field $A^a_\mu$ is quadratic in the framework of
GCM, one can perform the integration over gluon field
analytically. Using the same shorthand notation for the typical
Gaussian integrations as in Ref. \cite{r6}, we have
\begin{equation}\label{gvev}
\begin{split}
\int\mathcal{D}Ae^{-\frac{1}{2}AD^{-1}A+jA}
&={e}^{\frac{1}{2}jDj}\\
\int\mathcal{D}AAe^{-\frac{1}{2}AD^{-1}A+jA}
&=(jD){e}^{\frac{1}{2}jDj}\\
\int\mathcal{D}AA^2e^{-\frac{1}{2}AD^{-1}A+jA}
&=[D+{(jD)}^2]{e}^{\frac{1}{2}jDj}
\end{split}
\end{equation}
where $D$ is the dressing gluon propagator and $j_{\mu}^a$ is the
quark color current. Because the gluon vacuum average can be
replaced by a quark color current
$\overline{q}\gamma_\mu\frac{\lambda^a_C}{2}q$ together with the
gluon two-point function $D$, one can perform the integration over
the quark operators in the mean field vacuum as described above.
In this way, one can in principle obtain the vacuum expectation of
value for any gluonic fields. This technique provides an feasible
way to evaluate the expectation value of the operators with
low-dimensional gluon fields such as the mixed quark-gluon
condensate $g\langle\overline{q}G_{\mu\nu}\sigma^{\mu\nu}q\rangle$
in GCM (Note that for the VEV of operators with high powers of
gluonic fields, this procedure will get rather complex).
\par

\vspace{5pt} \noindent{\textbf{\Large{3. Mixed Tensor
Susceptibility}}} \vspace{5pt}

With above preparation, the mixed tensor susceptibility can be
calculated in the mean field level within this DSE model. The
induced tensor susceptibilities $\chi$, $\kappa$ and $\zeta$ are
defined through
\begin{eqnarray}
\langle{V}|\overline{q}\sigma_{\mu\nu}{q}|{V}\rangle_Z=g_q{\chi}Z_{\mu\nu}\langle\overline{q}q\rangle\\
\langle{V}|\overline{q}g_c\frac{\lambda^a}{2}G^a_{\mu\nu}{q}|{V}\rangle_Z=g_q{\kappa}Z_{\mu\nu}\langle\overline{q}q\rangle\\
\langle{V}|\overline{q}g_c\gamma_5\widetilde{G}_{\mu\nu}{q}|{V}\rangle_Z=-ig_q{\zeta}Z_{\mu\nu}\langle\overline{q}q\rangle,
\end{eqnarray}
where $Z_{\mu\nu}$ stands for the external field,
$\langle{V}|\cdots|{V}\rangle_Z$ denotes the VEV over the QCD
vacuum at the presence of the external filed $Z_{\mu\nu}$ and
$\widetilde{G}_{\mu\nu}=\frac{1}{2}\frac{\lambda^a}{2}\epsilon_{\mu\nu\alpha\beta}G^{\alpha\beta{a}}$.
The nonzero VEVs of the operators above are due to the breakdown
of Lorentz invariance in the presence of external constant field
$Z_{\mu\nu}$. From the QCD partition function for quarks in
Euclidean space in the presence of the external field , the
formulae for evaluating these susceptibilities take the form
\begin{eqnarray}\label{nsvev1}
\chi\langle\overline{q}{q}\rangle&=&\frac{1}{6}\int d^4x
\langle{V}|T[\overline{q}(x)\sigma_{\mu\nu}q(x),\overline{q}\sigma_{\mu\nu}q]|V\rangle=\frac{1}{6}\Pi_\chi(0)\\
\label{nsvev2}
\kappa\langle\overline{q}{q}\rangle&=&\frac{1}{6}\int d^4x
\langle{V}|T[\overline{q}(x)g_c\frac{\lambda^a}{2}G^a_{\mu\nu}q(x),\overline{q}\sigma_{\mu\nu}q]|V\rangle=\frac{1}{6}\Pi_\kappa(0),
\end{eqnarray}
according to Ref. \cite{r6,r7}. Due to the fact that the vacuum
susceptibilities reflect the nonperturbative structure of the QCD
vacuum, $\Pi_\chi(0)$ and $\Pi_\kappa(0)$ on the right hand side
of above Eqs.(\ref{nsvev1}) and (\ref{nsvev2}) should be
subtracted by the corresponding perturbative contribution terms.
Within the DSE formalism, the perturbative contribution to
$\Pi_\chi(0)$ and $\Pi_\kappa(0)$ can be evaluated by the trivial
quark propagator, namely the ``perturbative" quark propagator in
terms of the trivial Wigner solution to the dressed quark gap
equations (\ref{dsea}) and (\ref{dseb}). This is a reasonable
procedure because the Wigner solution of the dressed quark
propagator describes a phase with neither DCSB nor confinement and
the difference between the Nambu solution and the Wigner solution
vanishes at short distance according to numerical studies. In
fact, the Wigner solution can play the role of the perturbative
dressed quark propagator has been used extensively in the study of
thermal property of QCD within the DSE formalism\cite{r17}, where
the bag constant was defined as the difference of pressure between
the true QCD vacuum and the perturbative QCD vacuum, which were
evaluated by the Nambu-Goldstone solution and the Wigner solution
to the quark propagator, respectively\cite{r18}.
\par
Therefore, we rewrite the Eqs. (\ref{nsvev1}) and (\ref{nsvev2})
as
\begin{eqnarray}\label{svev1}
\chi\langle\overline{q}{q}\rangle&=&\frac{1}{6}\int d^4x
\langle{V}|T[\overline{q}(x)\sigma_{\mu\nu}q(x),\overline{q}\sigma_{\mu\nu}q]|V\rangle^N\nonumber\\
&&-\frac{1}{6}\int d^4x\langle{P}|T[\overline{q}(x)\sigma_{\mu\nu}q(x),\overline{q}\sigma_{\mu\nu}q]|P\rangle^W\nonumber\\
&=&\frac{1}{6}\Pi_\chi^{np}(0),\\
\label{svev2}\kappa\langle\overline{q}{q}\rangle&=&\frac{1}{6}\int
d^4x
\langle{V}|T[\overline{q}(x)g_c\frac{\lambda^a}{2}G^a_{\mu\nu}q(x),\overline{q}\sigma_{\mu\nu}q]|V\rangle^N\nonumber\\
&&-\frac{1}{6}\int d^4x\langle{P}|T[\overline{q}(x)g_c\frac{\lambda^a}{2}G^a_{\mu\nu}q(x),\overline{q}\sigma_{\mu\nu}q]|P\rangle^W\nonumber\\
&=&\frac{1}{6}\Pi_\kappa^{np}(0) .
\end{eqnarray}
By substituting the ``perturbative" quark propagator
$G^{per}(p^2)$ to Eq.(\ref{vev}), the determination of the
expectation value of the T-product operators in terms of quark
fields over the perturbative vacuum state $|P\rangle$ can be
performed self-consistently within the GCM formalism. It should be
noted that the evaluation of $\chi\langle\overline{q}{q}\rangle$
in Ref.\cite{r11} is consistent with Eq.(\ref{svev1}) because in
this special case the subtraction terms to Eq. (\ref{svev1})has
zero contribution to $\chi\langle\overline{q}{q}\rangle$ due to
$B'(p^2)\equiv{0}$ .
\par
Using Eq. (\ref{gvev}), the expression for VEV of above T-product
operator including gluonic fields is converted to the VEV for the
product with the form of (\ref{qop}) in terms of six quark fields
and eight quark fields. According to Eq. (\ref{vev}), we have

\begin{eqnarray}\label{mtsx}
\lefteqn{\frac{1}{6}\int d^4x
\langle{V}|T[\overline{q}(x)g_c\frac{\lambda^a}{2}G^a_{\mu\nu}q(x),\overline{q}\sigma_{\mu\nu}q]|V\rangle^N=}\nonumber\\
&&-\frac{4}{3}i\int dx^4\int
dz^4g^2\big[\partial_\mu^xD(z-x)\big]\times\nonumber\\
&&\bigg\{tr_D\big[G(x-z)\gamma_\upsilon
G(z-0)\sigma_{\mu\nu}G(0-x)\big]\nonumber\\
&&+tr_D\big[G(x-0)\sigma_{\mu\nu}G(0-z)\gamma_\upsilon
G(z-x)\big]\bigg\}\nonumber\\
&&-2i\int{dx^4}\int{dz_1^4}\int{dz_2^4}g^2D(z_1-x)g^2D(z_2-x)\times\nonumber\\
&&\bigg\{tr_D\big[G(x-z_1)\gamma_{\mu}G(z_1-z_2)\gamma_{\nu}G(z_2-0)\sigma_{\mu\nu}G(0-x)\big]\nonumber\\
&&+tr_D\big[G(x-z_1)\gamma_{\mu}G(z_1-0)\sigma_{\mu\nu}G(0-z_2)\gamma_{\nu}G(z_2-x)\big]\nonumber\\
&&+tr_D\big[G(x-0)\sigma_{\mu\nu}G(0-z_1)\gamma_{\mu}G(z_1-z_2)\gamma_{\nu}G(z_2-x)\big]\bigg\}.
\end{eqnarray}
Substituting $G^{per}(x-y)$ for $G(x-y)$, the similar expression
for the VEV of same operator over the perturbative vacuum can be
obtained. After Fourier transformation, we find the first part of
right hand side of (\ref{mtsx}) is zero and the final result for
the mixed tensor susceptibility in the momentum space takes the
form
\begin{eqnarray}\label{mts}
\kappa\langle\overline{q}{q}\rangle&=&\frac{1}{16\pi^2}\int ds
s[\frac{B(s)}{Z(s)}]^2[\frac{27}{8}B^2(s)+ \frac{27}{2}s
A(s)(2-A(s))]\nonumber\\
&&-\frac{9}{32\pi^5}\int dsdt\int_{-1}^{1} dx
st\sqrt{1-x^2}g^2D(s,t,\sqrt{st}x)Z^{-2}(s)Z^{-1}(t)\nonumber\\
&&\times
B(s)\bigg\{B(s)B^2(t)+B(t)A(s)A(t)\sqrt{st}x-\big[A(s)-1\big]\nonumber\\
&&\times \big[2A(s)B(t)\sqrt{st}x-A(t)B(s)\big]\bigg\}\nonumber\\
&&+\frac{3}{32\pi^5}\int dsdt\int_{-1}^{1} dx
st\sqrt{1-x^2}g^2D(s,t,\sqrt{st}x)Z'^{-2}(s)Z'^{-1}(t)\nonumber\\
&&\times A'^2(s)A'(t)\big[A'(t)-1\big]\big[st-4stx^2\big]\nonumber\\
&&-\frac{3}{32\pi^5}\int dsdt\int_{-1}^{1} dx
st\sqrt{1-x^2}g^2D(s,t,\sqrt{st}x)Z^{-2}(s)Z^{-1}(t)\nonumber\\
&&\times A^2(s)A(t)\big[A(t)-1\big]\big[st-4stx^2\big],
\end{eqnarray}
where $Z(s)=sA^2(s)+B(s)$ and $Z'(s)=sA'^2(s)$.
\par
It should be noted that to get this expression the Dyson-Schwinger
equation (\ref{DSE}) has been used again. The UV divergence of Eq.
(\ref{mtsx}) can be illustrated by a simple analytical confining
model
$g^2D(p-q)=\frac{3}{16}(2\pi)^4\eta^2\delta^4(p-q)$\cite{delta}.
In this model, the expression for (\ref{mtsx}) takes a relative
simple form
\begin{eqnarray}\label{delta}
\lefteqn{\frac{1}{6}\int d^4x
\langle{V}|T[\overline{q}(x)g_c\frac{\lambda^a}{2}G^a_{\mu\nu}q(x),\overline{q}\sigma_{\mu\nu}q]|V\rangle^N=}\nonumber\\
&&\frac{1}{16\pi^2}\int ds
s[\frac{B(s)}{Z(s)}]^2\big[\frac{27}{8}B^2(s)+ \frac{27}{2}s
A(s)(2-A(s))\big]\nonumber\\
&&-\frac{36\eta^2}{16\pi^2}\int ds s
Z^{-3}(s)\big[B^4(s)+sB^2(s)A(s)\big]\nonumber\\
&&+\frac{36\eta^2}{16\pi^2}\int ds s^3
Z^{-3}(s)\big[A(s)-1\big]A^3(s).
\end{eqnarray}
The Nambu-Goldstone solution for this model is
\begin{eqnarray}\label{NGA}
A(p^2)&=&\left\{
       \begin{array}{cc}
             2, &{ \quad p^2<\frac{\eta^2}{4}},\\
              \frac{1}{2}(1+\sqrt{1+\frac{2\eta^2}{p^2}}),
               &{\quad \mbox{otherwise}},
       \end{array}\right.\\
\label{NGB}B(p^2)&=&\left\{
       \begin{array}{cc}
             \sqrt{\eta^2-4p^2}, & {\quad p^2<\frac{\eta^2}{4}}, \\
            0, &{\quad
             \mbox{otherwise}}.
       \end{array}\right.
\end{eqnarray}
The alternative Wigner solution takes the form
\begin{equation}\label{Wig}
B'(p^2)\equiv0,\quad
A'(p^2)=\frac{1}{2}(1+\sqrt{1+\frac{2\eta^2}{p^2}}).
\end{equation}
Due to $A(s)-1 \sim c/s$  for $s\rightarrow \infty$ according to
(\ref{NGA}), the last term of right hand side of Eq. (\ref{delta})
is logarithmic divergent. In addition, replacing $A(s)$ with
$A'(s)$ to (\ref{delta}), there still exits logarithmic divergence
due to $A'(s)$ having the same behavior as $A(s)$ in the large
energy region. Because the vector part $A(s)$ as well as the
scalar part $B(s)$ both reflect the nonperturbative information in
the low energy region, it is more reasonable to subtract the
corresponding perturbative part rather than to simply ignore this
divergent term in (\ref{delta}).
\par
 With the effective subtraction
of the perturbative contribution, there is no UV divergence in
above integrations. Actually, the subtraction procedure guarantee
$\kappa\langle\overline{q}{q}\rangle$ the role of the order
parameter for QCD chiral phase transition because it becomes zero
when QCD undergoes a phase transition from the Nambu-Goldstone
phase to Wigner phase (That means $A(p^2)\rightarrow {A'(p^2)}$
and $B(p^2)\rightarrow{0}$ ).
\par
\vspace{5pt} \noindent{\textbf{\Large{4. Results and Discussions
}}} \vspace{5pt}
\par
The determination of the mixed tensor susceptibility is based on
the same effective gluon propagator models $g^2D(s)$
which had been used in Ref. \cite{r12,r19}. In general, the
quark-quark interaction $g^2D(s)$ has the form
\begin{equation}
g^2D(s)=\frac{4\pi\alpha(s)}{s},
\end{equation}
where $s=p^2$. Two popular quark-quark interaction models with two
parameters for $\alpha(s)$ are investigated here:
\begin{eqnarray}\label{int1}
\alpha_1(s)=3{\pi}s\chi^2\frac{e^{-\frac{s}{\Delta}}}{4\Delta^2}\\
\label{int2}\alpha_2(s)=d{\pi}s\frac{\chi^2}{s^2+\Delta},
\end{eqnarray}
where $d=\frac{27}{12}$. The two low-momentum parameters, the
stength parameter $\chi$ and the range parapmeter $\Delta$ , are
varied with the pion decay constant fixed at $87$ MeV which is
more appropriate in the chiral limit rather than the pion's
mass-shell value of $93$ MeV. Noted that the above quark-quark
interactions dominate for small $s$ and simulate the infrared
enhancement and confinement. Because the effective quark-quark
interactions(\ref{int1},\ref{int2}) have a finite range in
momentum space, the momentum integral for the calculation of the
quark condensate
\begin{equation}
\langle\overline{q}q\rangle=-\frac{3}{4\pi^2}\int_0^\infty{ds}s\frac{B(s)}{sA^2(s)+B^2(s)},
\end{equation}
is finite. According to \cite{r12}, the obtained values of the
chiral low energy coefficients $L_i$ following Ref.\cite{r19}
based on both ansatz (\ref{int1}) and (\ref{int2}) are compatible
with the phenomenological values. The model ansatz (\ref{int2})
has been successfully used to investigate the space structure of
the non-local quark condensate $\langle\overline{q}(x)q(0)\rangle$
in Ref.\cite{r20} within GCM formalism.
 It should be stressed  in this context that our
interactions are not renormalizable due to using the bare
quark-gluon vertex within the rainbow DSE formalism. Therefore,
the scale at which a condensate is defined in our calculation is a
typical hadronic scale, which is implicitly determined by the
model quark-quark interaction and the solutions of the DSEs for
the dressed quark propagator. The similar case is the
determination of the vacuum condensate in the instanton liquid
model\cite{r22} where the scale is set by the inverse instanton
size.

To check the sensitivity of the mixed tensor susceptibility on the
forms of quark-quark interactions, the above models with different
sets of parameters $\chi$ and $\Delta$ are investigated below,
where the results for the quark condensate are also given.
\par
\vspace{10pt}
\begin{tabular}{ccccc}
\multicolumn{5}{c}{Table 1. The value of
$\kappa\langle\overline{q}{q}\rangle$ for model 1 with
two sets }\\
\multicolumn{5}{c}{of different parameters.}
\\[2pt]\hline\hline
$\Delta(\mbox{GeV}^2)$&$\chi(\mbox{GeV})$&$-\langle\overline{q}{q}\rangle^{\frac{1}{3}}(\mbox{MeV})$
&$\kappa\langle\overline{q}{q}\rangle(\mbox{GeV}^4)$&$\kappa(\mbox{GeV})$
\\[2pt]\hline
0.2&1.55&213&$2.5*10^{-3}$&-0.26\\[2pt]
0.02&1.39&170&$2.9*10^{-3}$&-0.59\\[2pt]
0.002&1.30&149&$1.4*10^{-3}$&-0.41\\\hline\hline
\end{tabular}
\vspace{10pt}
\par
\begin{tabular}{ccccc}
\multicolumn{5}{c}{Table 2. The value of
$\kappa\langle\overline{q}{q}\rangle$ for model 2 with
four sets }\\
\multicolumn{5}{c}{of different parameters.}
\\[2pt]\hline\hline
$\Delta(\mbox{GeV}^4)$&$\chi(\mbox{GeV})$&$-\langle\overline{q}{q}\rangle^{\frac{1}{3}}(\mbox{MeV})$
&$\kappa\langle\overline{q}{q}\rangle(\mbox{GeV}^4)$&$\kappa(\mbox{GeV})$
\\[2pt]\hline
$1*10^{-1}$&1.77&290&$4.7*10^{-3}$&-0.19\\[2pt]
$1*10^{-2}$&1.33&250&$3.6*10^{-3}$&-0.23\\[2pt]
$1*10^{-4}$&0.95&217&$3.0*10^{-3}$&-0.29\\[2pt]
$1*10^{-6}$&0.77&204&$3.1*10^{-3}$&-0.36\\\hline\hline
\end{tabular}
\\[10pt]
\par
In Table 1 we display the values for $\langle\overline{q}q\rangle$
and $\kappa\langle\overline{q}q\rangle$ based on model 1 with
three sets of parameters and in Table 2 the same quantities with
four sets of parameters based on model 2. In both cases, the
obtained values for $\langle\overline{q}q\rangle$ are compatible
with the standard phenomenological value in  SVZ sum rules,
whereas the mixed tensor susceptibility
$\kappa\langle\overline{q}q\rangle$ is much small. The previous
estimation of $\kappa\langle\overline{q}q\rangle$ obtained In Ref.
\cite{r6} is $0.10$ $\mbox{GeV}^4$. Actually, the value of the
quark condensate tensor susceptibility
$\chi\langle\overline{q}q\rangle$ obtained within GCM formalism
\cite{r11} is also very small compared with the estimation based
on SVZ sum rules . In fact, different versions of SVZ sum rules
have given very different values for
$\chi\langle\overline{q}q\rangle$ in previous studies
\cite{r6,r8,r9}.  Therefore, it shows that the induced vacuum
condensates have very little impact on the determination of the
nucleon tensor charge from the theoretical formalism of DSEs.
\par
In summary, we have investigated the mixed tensor susceptibility
at the mean field level in the framework of GCM/DSE formalism. In
the calculations, the vacuum matrix elements for the operator in
terms of quark and gluonic fields can be obtained by substituting
the gluonic fields with the quark color current operator and the
model gluon propagator which describes the effective quark-quark
interaction within the GCM formalism. To subtract the perturbative
contribution to the expression for the mixed tensor
susceptibility, the Wigner solution to the quark gap equation was
used self-consistently in this formalism. Using different
quark-quark interaction models, we find that the mixed tensor
susceptibility  as well as the quark condensate tensor are both
very small susceptibility does within DSE formalism. Therefore, we
get the conclusion that the induced vacuum condensates have little
effect on the determination of the nucleon tensor charge. Finally,
we want to tress that above approach can also be used to
investigate the other mixed susceptibility of the QCD vacuum.
\\[10pt]
\noindent{\textbf{\Large{Acknowledgements}}} \vspace{10pt}
\par
We would like to thank Jia-lun Ping for his help to the numerical
study.

\end{document}